\documentclass[aps,prd,twocolumn,floatfix,showpacs,superscriptaddress,nofootinbib]{revtex4-2}
\usepackage{amsmath,amsfonts,amssymb,bm}
\usepackage{diagbox}
\usepackage{graphicx}
\usepackage{xcolor}
\usepackage{subfigure}
\usepackage{multirow}
\usepackage{textcomp}
\usepackage{slashed}
\usepackage[bottom]{footmisc}
\usepackage[T1]{fontenc}
\definecolor{purple}{rgb}{0.5,0,0.5}
\definecolor{blue}{rgb}{0.0,0,1.0}
\usepackage[colorlinks=true, pdfstartview=FitV, linkcolor=purple, citecolor= purple, urlcolor=blue]{hyperref}

\newcommand{\ga}{\gamma}

\newcommand{\beq}{\begin{equation}}
\newcommand{\eeq}{\end{equation}}
\newcommand{\ba}{\begin{array}}
\newcommand{\ea}{\end{array}}
\newcommand{\bea}{\begin{align}}
\newcommand{\eea}{\end{align}}
\newcommand{\bi}{\begin{itemize}}
\newcommand{\ei}{\end{itemize}}
\newcommand{\ben}{\begin{enumerate}}
\newcommand{\een}{\end{enumerate}}
\newcommand{\bc}{\begin{center}}
\newcommand{\ec}{\end{center}}
\newcommand{\bl}{\begin{flushleft}}
\newcommand{\el}{\end{flushleft}}
\newcommand{\br}{\begin{flushright}}
\newcommand{\er}{\end{flushright}}

\begin{document}



\title{Radiative charmonium decays in a contact-interaction model with dynamical quark anomalous magnetic moment}

\author{Yehan Xu}
\affiliation{School of Physics, Nankai University, Tianjin 300071, China}

\author{Zanbin Xing}
\affiliation{School of Physics, Nankai University, Tianjin 300071, China}

\author{Kh\'epani Raya}
\affiliation{Department of Integrated Sciences and Center for Advanced Studies in Physics, Mathematics and Computation, University of Huelva, E-21071 Huelva, Spain.}

\author{Lei Chang}
\affiliation{School of Physics, Nankai University, Tianjin 300071, China}

\date{\today}
\begin{abstract}
The BESIII Collaboration has recently reported two measurements of the two-photon decay width of the $\eta_c$ meson. The 2024 result is significantly larger than most theoretical and empirical expectations, while a subsequent measurement published in early 2026 shows better agreement with the world average and conventional theoretical estimates. In this work, we study the $\eta_c\to\gamma\gamma$ and $J/\psi\to\gamma\eta_c$ processes within a contact interaction model that incorporates valence-quark anomalous magnetic moment effects, which are absent in standard treatments. Besides achieving agreement with modern lattice QCD estimates for these observables, we find that the 2024 central value for $\eta_c\to\gamma\gamma$ lies above the range that could be accommodated by the present framework, whereas the 2026 result is naturally consistent with it.
\end{abstract}

\maketitle
\section{Introduction}\label{sec:intro}

Two-photon couplings of pseudoscalar mesons provide a unique window into different facets of quantum chromodynamics (QCD): from anomaly-driven processes in the light sector,\,\cite{Adler:1969gk,Bell:1969ts,Feldmann:1997vc,Agaev:2014wna}, governed by dynamical chiral symmetry breaking (DCSB),\,\cite{Raya:2024ejx}, to short-distance dominated processes in heavy quarkonia, where factorization and perturbative methods become applicable,\,\cite{Feldmann:1997te,Feng:2015uha}. In this context, the two-photon decay of the $\eta_c$ serves as a benchmark for charmonium dynamics and a sensitive probe of QCD. Over the years, the decay width $\Gamma_{\eta_c\to\gamma\gamma}$ has been measured by many groups, including two recent (and seemingly contradictory) determinations by the BESIII Collaboration in 2024\,\cite{BESIII:2024rex} and 2026\,\cite{BESIII:2026pff}. The former reports a value that exceeds the world average, $\Gamma_{\eta_c\to\gamma\gamma}^{\text{PDG}}=5.1\pm0.4\,\text{keV}$,\,\cite{ParticleDataGroup:2024cfk}, by more than $3\sigma$. Most theoretical predictions also disfavor the 2024 determination, including: lattice QCD\,\cite{Liu:2020qfz,Meng:2021ecs,Colquhoun:2023zbc}, non-relativistic QCD\,\cite{Feng:2015uha,Feng:2017hlu,Li:2019ncs,Wang:2025afy}, the light-front Hamiltonian approach\,\cite{Li:2021ejv}, and continuum Schwinger methods (CSMs)\,\cite{Chen:2016bpj,Raya:2016yuj,Raya:2019dnh}. By contrast, the 2026 BESIII result lies closer to conventional expectations. The tension between these two BESIII measurements, obtained within the same collaboration, complicates a straightforward interpretation and underscores the need for further clarification before the $\eta_c \to \gamma\gamma$ width can be regarded as a settled observable.

In light of the current experimental situation, it is timely to revisit the $\eta_c\to\gamma\gamma$ decay, together with the associated $J/\psi\to\gamma\eta_c$ process, from a theoretical standpoint. In particular, one should assess whether relevant physical effects have been overlooked or only partially accounted for in conventional approaches. The $\eta_c$ meson exhibits a nontrivial internal structure shaped by the nonperturbative dynamics of QCD. Among these effects is the emergence of a valence-quark anomalous magnetic moment (AMM). For light quarks, it is sizable and comparable in magnitude to the anomalous chromomagnetic moment~\cite{Chang:2010hb,Bashir:2011dp}, and may also induce non-negligible contributions in the charm sector. The strength of the AMM is, in fact, related with that of DCSB. Within the $\eta_c$, the AMM of its valence-constituents can be interpreted as an effective \emph{magnetic dipole} that directly influences its interaction with photons~\cite{Chang:2010hb}.  While previous studies have often neglected this effect or treated it within a mostly perturbative perspective~\cite{Feng:2017hlu}, it remains worthy of a dedicated analysis. Lying at the interface between the DCSB (soft physics) and perturbative domains, the charm quark provides a natural setting to investigate the transition between these regimes, and so the $\eta_c \to \gamma\gamma$ decay offers an ideal platform to explore how AMM effects interpolate between soft and hard dynamics.

In principle, as discussed in Ref.\,\cite{Chang:2010hb,Bashir:2011dp}, the quark AMM can be incorporated in a rigorous and self-consistent manner within CSMs. Over the years, this approach has led to a deeper understanding of the structure of the quark propagator and its implications for hadron spectroscopy; see, e.g.\, Refs.~\cite{Chang:2021vvx,Chang:2011ei,Qin:2020jig}. However, while detailed analyses of $\gamma^*\gamma^{(*)}\to$ neutral pseudoscalar transitions exist, e.g.\, Refs.~\cite{Chen:2016bpj,Raya:2016yuj,Raya:2019dnh}, these do not allow for a systematic incorporation or a clear disentanglement of AMM contributions. As a first step toward filling this gap, we examine the $\eta_c\to\gamma\gamma$ and $J/\psi\to\gamma\eta_c$ processes within the symmetry-preserving CSM scheme developed in Ref.~\cite{Xing:2021dwe}. In particular, we employ an illustrative contact interaction (CI) model that incorporates AMM effects in a phenomenological yet symmetry-consistent manner. This contrasts with earlier studies of two-photon transitions to heavy quarkonia~\cite{Bedolla:2016yxq,Raya:2017ggu}, where AMM effects were neglected and the corresponding decay widths were not reported.

It is worth noting that the CI model is not intended for high-precision quantitative predictions. As demonstrated in studies of various anomalous processes,\,\cite{Dang:2023ysl,Xing:2024bpj,Xu:2024frc,Xing:2025iab}, its main utility lies in preserving key symmetries of QCD and its non-perturbative character, while offering a transparent and controllable framework to explore the qualitative impact of dynamical mechanisms —such as the AMM effect— on hadronic observables. Accordingly, our analysis does not aim to \emph{rule out} or \emph{validate} any experimental measurement in an absolute sense. Instead, we evaluate whether a given measurement can be accommodated within this symmetry-preserving, self-consistent framework under reasonable parameter variations, particularly those associated with the AMM contributions. A lack of compatibility would indicate either the need for additional dynamical contributions beyond the AMM effect or motivate further experimental scrutiny; it should not be interpreted as a statement on the correctness of the measurement itself. With this caveat in mind, we evaluate the impact of the AMM effect within the CI model. Even after its inclusion, the predicted decay widths remain below the 2024 BESIII central value across reasonable parameter choices, while being compatible with the 2026 measurement. This points to a tension that warrants further clarification through independent experimental input and more refined theoretical analyses. 

The manuscript is organized as follows. In Sec.~\ref{sec:formalism}, we present the CI model formalism, including the relevant one- and two-body equations, the corresponding diagrams, as well as  the parameter-fixing procedure. The resulting masses and decay constants are presented in Sec.~\ref{sec:parameters}, while~\ref{sec:results} is devoted to the evaluation of decay widths. Finally, a brief summary is given in Sec.~\ref{sec:summary}.

\section{FORMALISM}\label{sec:formalism}
\subsection{One- and Two-body equations}\label{sec:CI}

The CI model is characterized by the following construction for the gluon propagator $D_{\mu\nu}$:
\begin{equation}
g^2D_{\mu\nu}(p-q)=\delta_{\mu\nu}\frac1{m_G^2}\,,
\end{equation}
together with a tree-level structure for the corresponding quark-gluon vertex $\Gamma^{a}_{\nu}$:
\begin{equation}
\Gamma_{\nu}^a(q,p)=\frac{\lambda^a}{2}\gamma_{\nu}\,.
\end{equation}
Here $m_G$ plays the role of an effective gluon mass scale, which is believed to emerge in QCD as a manifestation of the Schwinger mechanism\,\cite{Ferreira:2025tzo,Aguilar:2010gm}. 

The $f$-flavored dressed-quark propagator, $S_f$, can be expressed as:
\begin{equation}
S_{f}^{-1}(p)=i\gamma\cdot p+M_f\,,\end{equation}
where the momentum-independent quantity $M_f$ plays the role of a constituent quark mass and is determined from the corresponding gap equation:
\begin{equation}\label{eq:Mf}
M_{f}=m_{f}+\frac{16}{3m_G^2}\int_q\frac{M_f}{q^2+M_f^2}\,,
\end{equation}
with $m_f$ denoting the current quark mass, and $\int_q\doteq\frac{d^4q}{(2\pi)^4}$ representing a Poincar\'{e} invariant integration.

Note that the integral in Eq.\,\eqref{eq:Mf} is divergent and, in general, the CI yields both logarithmic and quadratic divergences. The symmetry-preserving regularization scheme developed in Ref.\,~\cite{Xing:2022jtt} allows one to handle these divergences by classifying the resulting integrals into two types:
\begin{align}
\int_{q}\frac{1}{(q^{2}+\mathcal{M}^{2})^{\alpha+2}}
&\overset{R}\to\,I_{-2\alpha}^R(\mathcal{M}^{2})\,,\label{eq:Re}\\
\int_q\frac{q_{\mu}q_{\nu}}{(q^2+\mathcal{M}^2)^{\alpha+3}}
&\overset{R}\to\,\frac{\Gamma(\alpha+2)}{2\Gamma(\alpha+3)}\delta_{\mu\nu}I_{-2\alpha}^R(\mathcal{M}^{2})\,,
\end{align}
where we have defined:
\begin{equation}
    I_{-2\alpha}^R(\mathcal{M}^{2}):=\int_{\tau_{uv}^{2}}^{\tau_{ir}^{2}}d\tau\frac{\tau^{\alpha-1}}{\Gamma(\alpha+2)}\frac{e^{-\tau\mathcal{M}^{2}}}{16\pi^{2}}\,.
\end{equation}
The quantities $\tau_{ir}=1/\Lambda_{ir}$ and $\tau_{uv}=1/\Lambda_{uv}$ denote the infrared (IR) and ultraviolet (UV) cutoffs, respectively. The former is of the order of $\Lambda_{ir}\sim\Lambda_{\text{QCD}}$ and yields a confinement-compatible description by eliminating quark production thresholds. The ultraviolet scale $\Lambda_{uv}$ plays a dynamical role and cannot be removed. Its preferred range of values is discussed below.

Our description of quark–antiquark bound states, as well as the quark–photon vertex (QPV), extends that of the conventional rainbow-ladder (RL) truncation in the CI model, e.g. Refs.\,\cite{Gutierrez-Guerrero:2010waf,Roberts:2010rn}. Here we adopt the modified rainbow–ladder (MRL) truncation introduced in Ref.~\cite{Xing:2021dwe}. Among its desirable features, this scheme generates an AMM term in a self-consistent and fully dynamical manner. The Bethe–Salpeter equations (BSEs) for the $H$-meson amplitude, $\Gamma_H$, and the quark–photon vertex, $\Gamma_\mu$, are given by:
\begin{eqnarray}
    \label{eq:CIinMRL}
\Gamma_H(P)=-\frac{4}{3m_{G}^{2}}\int_{q}\left[\gamma_{\alpha}\chi_H(P)\gamma_{\alpha} - \xi\,\tilde{\Gamma}_{j}\chi_H(P)\tilde{\Gamma}_{j}\right]\,,\hspace{0.1cm}\\
\label{eq:QPVinMRL}
\Gamma_\mu(Q)=\gamma_\mu-\frac{4}{3m_{G}^{2}}\int_{q}\left[\gamma_{\alpha}\chi_\mu(Q)\gamma_{\alpha} - \xi\, \tilde{\Gamma}_{j}\chi_\mu(P)\tilde{\Gamma}_{j}\right]\,,\hspace{0.1cm}
\end{eqnarray}
where $P$ and $Q$ denote the meson and photon momenta, respectively. Owing to the CI prescriptions, neither amplitude depends on the relative momentum between the quark and antiquark. The Bethe–Salpeter wave function reads $\chi(q;P)=S(q)\Gamma(q;P)S(q-P)$, and the un-amputated QPV $\chi_\mu(q;P)$ is defined analogously. The beyond rainbow-ladder component is defined by $\tilde{\Gamma}{j} = \{ \mathbb{I},\;\gamma_5,\;i\sigma_{\mu\nu}/\sqrt{6}\}$, with the dimensionless parameter $\xi$ controlling its strength. This term encodes the dynamically generated AMM, which is expected to be flavor dependent.

The pseudoscalar ($\mathcal{P}$) meson BSA, vector ($\mathcal{V}$) meson BSA, and QPV can be decomposed into the following general structures:
\begin{eqnarray}
\label{eq:BSAps}
\Gamma_{\mathcal{P}}(P) = i \gamma_5 E_{\mathcal{P}}(P) + \frac{1}{\bar{M}}\gamma_5\gamma \cdot P F_{\mathcal{P}}(P) \;,\hspace{1.9cm}\\
\label{eq:BSAv}
\Gamma^{\mathcal{V}}_{\mu}(P)=\gamma_{\mu}^{T}E_{\mathcal{V}}(P)+\frac{1}{\bar{M}}\sigma_{\mu\nu}P_{\nu} F_{\mathcal{V}}(P)\;,\hspace{2.15cm}\\
\label{eq:qpvstructure}
\Gamma_\mu(Q)=\gamma_\mu^Lf_L(Q^2)+\gamma_\mu^Tf_T(Q^2)+\frac{1}{\bar{M}}\sigma_{\mu\nu}Q_{\nu}f_A(Q^2)\,,\hspace{0.1cm}
\end{eqnarray}
with $\gamma_{\mu}^{T}(P)=\gamma_{\mu}-\frac{ \gamma \cdot P \;
 }{P^{2}} \, P_{\mu}$, $\gamma_\mu^L=\gamma_\mu-\gamma_\mu^T$; and $\bar{M}=2M_fM_h/(M_q+M_h)$, where $f$ and $h$ denote the valence quark and antiquark flavors in the meson $H$. For quarkonia systems, $\bar{M}=M_f$.
 
Being solutions of a homogeneous BSE, the BSAs must satisfy the canonical normalization condition in order to yield physical observables. This condition reads:
\begin{equation}\begin{aligned}2P_\mu=\mathrm{tr}\int_q\Gamma_H(-P)S(q)\Gamma_H(P)\frac{\partial}{\partial P_\mu}S(q-P)\end{aligned}\,,\end{equation}
where the trace is taken over color, flavor, and spinor indices. The leptonic decay constants of pseudoscalar and vector mesons, respectively, can be then obtained via the following expressions:
\begin{eqnarray}\sqrt{2}\,P_\mu f_{\mathcal{P}}&=&\mathrm{tr}\int_q\left[\gamma_5\gamma_\mu S(q)\Gamma_{PS} S(q-P)\right]\,,\label{eq:decaycPS}\\
\sqrt{2}\,m_\mathcal{V}f_\mathcal{V}&=&\frac{1}{3}\delta_{\mu\nu}\mathrm{tr}\int_q\left[\gamma_\mu S(q)\Gamma_\nu^V S(q-P)\right]\,.
\end{eqnarray}
With the relevant propagators and vertices in hand, together with the properly normalized BSAs, we now proceed to compute the radiative decays and form factors. This is described below.

\subsection{Radiative Decays in the Impulse Approximation}

The matrix elements describing the processes $\mathcal{P} \to \gamma^*\gamma$ and $\mathcal{V}\rightarrow\gamma^*\mathcal{P}$ read, respectively:
\begin{equation}\begin{aligned}
\label{eq:mathelem}
&T_{\mu\nu}^{\mathcal{P}\gamma}(k_1;k_2)=e^2g_{\mathcal{P}\gamma}\epsilon_{\mu\nu\rho\sigma}k_1^{\rho}k_2^{\sigma}\,G_{\mathcal{P}\gamma}\left(k_1,k_2\right)\,,
\\&T_{\mu\nu}^{\mathcal{V}\mathcal{P}}(k_3;k_4)=e\,g_{\mathcal{VP}}\,\epsilon_{\mu\nu\rho\sigma}\,k_3^{\rho}k_4^{\sigma}F_{\mathcal{V}\mathcal{P}}\left(k_3,k_4\right) \,.
\end{aligned}\end{equation}
Herein, $k_{1}$ and $k_{3}$ denote the momenta of the off-shell photon ($\gamma^*$) in each process, while $k_{2}$ corresponds to the on-shell photon ($\gamma$) in the two-photon transition and $k_4$ to the pseudoscalar meson momentum in $\mathcal{V}\rightarrow\gamma^*\mathcal{P}$. The couplings $g_{\mathcal{P}\gamma}$ and $g_{\mathcal{VP}}$ are defined such that the associated form factors are normalized as $G_{\mathcal{P}\gamma}(0,0)=1$ and $F_{\mathcal{V}\mathcal{P}}(0,0)=1$. With these definitions, the corresponding decay widths are given by:
\begin{equation}
\begin{aligned}
\label{eq:decformg}
    &\Gamma_{\mathcal{P}\gamma} = \frac{1}{4}\pi \alpha_{em}^2 \,g_{\mathcal{P}\gamma}^2\,m_\mathcal{P}^3\,,
\\
&{\Gamma_{\mathcal{V}\mathcal{P}}=\frac{1}{3}\alpha_{\mathrm{em}}\,g_{\mathcal{V}\mathcal{P}}^2\,{\left(\frac{m_{\mathcal{V}}^2-m_{\mathcal{P}}^2}{2m_{\mathcal{V}}}\right)^3}}.
\end{aligned}
\end{equation}

Both processes admit a description within the impulse approximation, where the external probe couples to a single valence constituent and the rest act as spectators. At a truncation level consistent with the present one- and two-body framework, this leads to:
\begin{equation}
\begin{aligned}
\label{eq:triang}
T_{\mu\nu}^{\mathcal{P}\gamma}(k_1;k_2)&=2 \,\text{tr}\int_{q}\big[S(q)i\Gamma_{\mu}(-k_1)S(q+k_1)\\&\times\Gamma^{\mathcal{P}}(k_1+k_2)S(q-k_2)i\Gamma_{\nu}(-k_2)\big]\,,
\\
T_{\mu\nu}^{\mathcal{V}\mathcal{P}}(k_3;k_4)&=2 \,\text{tr}\int_{q}\big[S(q)\Gamma^{\mathcal{P}}(-k_3)S(q+k_3)\\
&\times\Gamma_{\mu}^{\mathcal{V}}(k_3+k_4)S(q-k_4)i\Gamma_{\nu}(-k_4)\big]\,,\end{aligned}
\end{equation}
where the trace is taken over the color and spinor
indices, and the overall factor of $2$ in each expression originates from the fact that we consider quarkonia systems. The on-shell meson conditions entail $(k_1+k_2)^2=-m_{\mathcal{P}}^2=k_3^2,\,(k_3+k_4)^2=-m_{\mathcal{V}}^2$. 

It is worth noting that regularization subtleties may affect results involving an odd number of $\gamma_5$ matrices. In line with previous analyses of anomalous processes within the CI model~\cite{Dang:2023ysl,Xing:2024bpj,Xu:2024frc,Xing:2025iab}, we adopt the representation:
\begin{equation}
    \gamma_5=-\frac{1}{24}\epsilon_{abcd}\gamma_{a}\gamma_{b}\gamma_{c}\gamma_{d}\,.
\end{equation}
The remaining task is to determine the free parameters that define the CI in the charmonium sector.

\subsection{Parameter Determination}\label{sec:parameters}

Since the earliest studies of heavy quarkonia in the CI model, Refs.\,\cite{Raya:2017ggu,Bedolla:2016yxq}, it has been suggested that the ultraviolet cutoff should depend on the mass sector under consideration; in particular, for systems with heavier quarks, $\Lambda_{uv}$ should exceed the standard light sector value $\Lambda_{uv}\sim 0.9\,\text{GeV}$. This extended momentum range corresponds to a more localized structure in coordinate space, as expected for heavier systems\,\cite{Raya:2024ejx}. Typical charmonium scales correspond to $\Lambda_{uv}\gtrsim 2\,\text{GeV}$, e.g.\,\cite{Yin:2021uom,Gutierrez-Guerrero:2021rsx}; accordingly, we explore the range $\Lambda_{uv}=2-2.5\,\text{GeV}$. 

The role of the infrared cutoff, as noted previously, is to prevent quark production thresholds, and its standard value, $\Lambda_{ir}=0.24\,\text{GeV}\sim \Lambda_{\text{QCD}}$, does not require modification. Similarly, the chiral anomaly fixes $\xi=0.151$,\,\cite{Dang:2023ysl}. The range $\xi = 0 - 0.151$ shall be explored later in order to better elucidate the role of the AMM. As far as the meson spectrum is concerned, the choice of $\xi$ would only affect the vector channel\,\cite{Xing:2021dwe}.

The remaining mass-dimensioned parameters, $m_c$ and $m_G$, are fixed by reproducing the masses of the ground-state pseudoscalar and vector mesons, namely $\eta_c$ and $J/\psi$. Together with $\Lambda_{uv}$, these constitute the only free parameters in our scheme. Their values, along with the resulting constituent masses, are listed in Table\,\ref{tab:paramsGap}; the corresponding meson masses and leptonic decay constants are reported in Table\,\ref{tab:meson}. Clearly, the computed $m_{\eta_c}$ and $m_{J/\psi}$ faithfully reproduce the experimental values reported by the Particle Data Group (PDG)\,\cite{ParticleDataGroup:2024cfk}, and the associated leptonic decay constants compare favorably with lattice QCD (lQCD) results\,\cite{McNeile:2012qf,Donald:2012ga}. 

\begin{table}[!h]
\renewcommand{\arraystretch}{1.2}
\begin{tabular}{c c c|c}
\hline
     $\Lambda_{uv}$ & $m_G$ &  $m_c$ & $M_{c}$\\
    \hline 
     $2$   & $0.4048$ &$1.1927$ & $1.54613$ \\

     $2.5$ & $0.5476$ &$1.0578$ & $1.50008$ \\
\hline
\end{tabular}
\caption{\label{tab:paramsGap}(left) Lower and upper values of $\Lambda_{uv}$ and the correponding parameters $m_G$ and $m_c$. (right) The resulting constituent mass $M_c$. In all cases, $\Lambda_{ir}=0.24\,\text{GeV}$ and $\xi=0.151$. Mass units in GeV.}
\end{table}

\begin{table}[!h]
\renewcommand{\arraystretch}{1.2}
\begin{tabular}{c|c c||c|c c}
\hline
     & $\text{MRL}_{avg.}$ & $\text{PDG}$ &&$\text{MRL}_{avg.}$ &$\text{lQCD}$ \\
    \hline 
     $m_{\eta_c}$ & $2.98408(7)$ & $2.9841(4)$ &$f_{\eta_c}$& $0.261(10)$ & $0.278(2)$ \\
     $m_{J/\psi}$ & $3.09687(4)$ & $3.096900(6)$ &$f_{J/\psi}$& $0.212(2)$ & $0.286(4)$  \\
\hline
\end{tabular}
\caption{\label{tab:meson}Average values and standard errors of meson masses and leptonic decay constants obtained using the parameter ranges in Table\,\ref{tab:paramsGap}. Experimental masses from PDG,\,\cite{ParticleDataGroup:2024cfk}, and decay constants from lQCD,\,\cite{McNeile:2012qf,Donald:2012ga}. Mass units in GeV.}
\end{table}

\section{RESULTS AND DISCUSSION}\label{sec:results}

Let us start this discussion by highlighting key aspects of QPV in the present framework. According to Eq.\,\eqref{eq:qpvstructure}, the vertex decomposes into a longitudinal component $f_L$, and two transverse components $f_T$ and $f_A$. The last component arises dynamically within the MRL truncation, respects all relevant symmetries, and vanishes in the limit $\xi\to0$. Combined with Eq.\,\eqref{eq:QPVinMRL}, one finds that $f_L(Q^2)=1$, thus ensuring the preservation of the longitudinal Ward-Green-Takahashi identity,\,\cite{Ward:1950xp,Green:1953te, Takahashi:1957xn}.

In the ultraviolet limit, $Q^2\to\infty$, the asymptotic behavior of the two transverse components are:
\begin{equation}
    f_T(Q^2\to\infty)\to 1\,,\,f_A(Q^2\to\infty)\to 0\,,
\end{equation}
such that the QPV recovers the point-particle profile $\Gamma_\mu(Q^2\to\infty)\to \gamma_\mu$ due to the vanishing of non-perturbative dressing effects in this limit. In the soft regime $Q^2\approx0$, we find $f_T(0)=1$ and $f_A(0)\neq 0$ for $\xi\neq0$. The $f_A$ term thus induces a dynamical enhancement of the bare vertex $\gamma_\mu$, reflecting the DCSB-driven contributions associated with the AMM. The soft-limit values of the QPV dressing functions, together with their slopes
\begin{equation}
\label{eq:defSlope}
    R_{T,A}:=-\frac{\partial f_{T,A}(Q^2)}{\partial Q^2}|_{Q^2=0}\,,
\end{equation}
are summarized in Table\,\ref{tab:AMMval} for both $u/d$ and charm quarks. The AMM correction for the $c$ quark is significantly smaller than that for the light $u/d$ quarks, amounting to roughly one half, $1/2$. Likewise, the slopes of the dressing functions are less pronounced. These findings are consistent with expectations: as the quark mass increases, non-perturbative (DCSB-driven) effects are suppressed, so that quarks behave increasingly like point-like particles, reflected in a reduced AMM contribution and a less pronounced $Q^2$ dependence of the QPV. 
\begin{table}[!htbp]
\renewcommand{\arraystretch}{1.2}
    \centering
        
    \begin{tabular}{c c c c c }
        \hline
         & $f_{T}^{f}$ & $f_{A}^{f}$ &  $R_{T}^{f}$ &  $R_{A}^{f}$  \\
        \hline
         $u/d$ & $1.0$ & $0.043$ & $0.4933$& $0.0556$ \\
         $c_{avg.}$& $1.0$&  $0.023(1)$&  $0.0161(5)$& $0.0022(3)$   \\
        \hline
    \end{tabular}
    \caption{\label{tab:AMMval}QPV dressing functions and their slope(in GeV) at $Q^2=0$. The results for c-quark are the average value, with the standard error listed. Relevant data on light quarks is cited from Ref.\,\cite{Xu:2024frc}, with the parameter $\Lambda_{uv}=0.905$.}
    \label{tab:QPV}
\end{table}

Having discussed essential aspects of the quark-photon interaction, we now turn to the evaluation of the decay widths and coupling constants associated with the processes $\eta_c\to\ga^*\ga$ and $J/\psi\to\ga^*\eta_c$, defined in Eq.\,\eqref{eq:mathelem}. The computed results for RL and MRL truncation are listed in Table\,\ref{tab:Decays}. The first row reports the averaged RL results, corresponding to $\xi=0$, while the second row presents the MRL expectations. In the latter, we consider not only the preferred value $\xi=0.151$ (see Table\,\ref{tab:paramsGap}), but also its variation over the range $\xi=0-1.151$\,\footnote{Note that for every value $\Lambda_{uv}$ and $\xi$, $m_G$ and $m_c$ are fitted to produce the experimental values $m_{\eta_c}$ and $m_{J/\psi}$}. Recent lQCD results from Refs.\,\cite{Meng:2024axn,Colquhoun:2023zbc} are also shown for comparison. Both RL and MRL scenarios show compatibility with lattice QCD. However, the $J/\psi\to\ga^*\eta_c$ case is better described by the MRL truncation. Moreover, both the decay widths and the couplings associated to $\eta_c\to\ga^*\ga$ and $J/\psi\to\ga^*\eta_c$ are enhanced in the MRL case, highlighting the contribution of the AMM to the electromagnetic interaction. The corresponding uncertainties are visibly larger, as the full variation range of $\xi$ is considered, further emphasizing the influence of the AMM. The $Q^2$ dependence of these transitions is briefly discussed in Appendix\,\ref{ap:TFFs}.

\begin{table}[!htbp]
\renewcommand{\arraystretch}{1.2}
    \centering
          
    \begin{tabular}{c|c c c c}
        \hline
        & $|g_{\mathcal{P}\gamma}|$ & $|g_{\mathcal{V}\mathcal{P}}|$ & $\Gamma_{\mathcal{P}\gamma}$ & $\Gamma_{\mathcal{V}\mathcal{P}}$   \\
        \hline
         $\text{RL}_{avg.}$ & $0.075(1)$ & $0.780(6)$ & $6.23(18)$ & $2.01(3)$\\
         $\text{MRL}_{avg.}$ & $0.082(6)$ & $0.798(17)$ & $7.48(114)$ & $2.11(9)$\\
         
        \hline
        
        $\text{lQCD}$\,\cite{Colquhoun:2023zbc} & $0.078(1)$ & $0.819(6)$ & $6.79(6)$ & $2.22(3)$\\
        $\text{lQCD}$\,\cite{Meng:2024axn} & $0.077(10)$ & $0.834(18)$ & $6.67(17)$ & $2.30(10)$\\
        \hline
    \end{tabular}
    \caption{Decay widths $\Gamma_{\mathcal{P}\gamma}$ and $\Gamma_{\mathcal{V}\mathcal{P}}$ (in keV), together with the corresponding coupling constants $|g_{\mathcal{P}\gamma}|$ and $|g_{\mathcal{V}\mathcal{P}}|$ (in GeV$^{-1}$), for the charmonium processes $\eta_c\to\ga^*\ga$ and $J/\psi\to\ga^*\eta_c$. The errors in the lQCD evaluation from Ref.\,\cite{Colquhoun:2023zbc} have been combined in quadrature.}
    \label{tab:Decays}
    \end{table}

\begin{figure}[!htbp]
 \centering
 \includegraphics[scale=0.69]{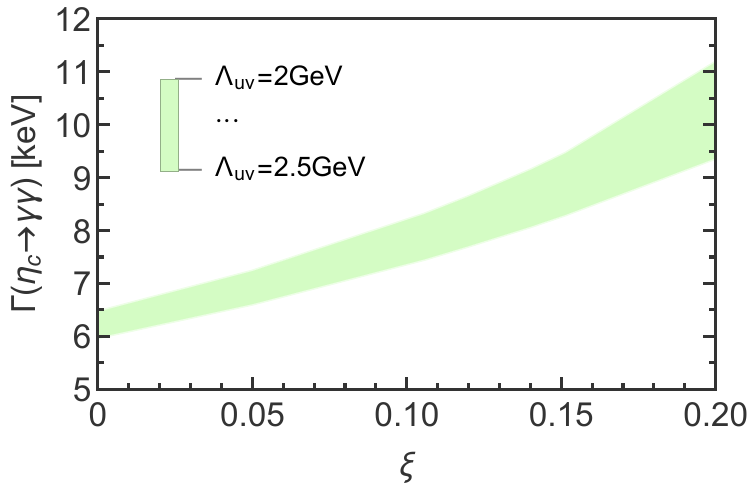}
   \caption{The $\xi$-dependence of the $\Gamma[\eta_c\to\gamma\gamma]$ decay width. The band denotes the uncertainty associated with the variation of $\Lambda_{uv}$, where the upper (lower) bound corresponds to $\Lambda_{uv}=2\,\text{GeV}\,(2.5\,\text{GeV})$.}
   \label{fig:xi}
\end{figure}

Regarding the variation of $\xi$, it is important to recall that this parameter weights contributions beyond the RL truncation, which are responsible for the emergence of the AMM term. Since these contributions are driven by the strength of DCSB, they are expected to be flavor dependent and to have a stronger impact in the light-quark sector\,\cite{Bashir:2011dp,Chang:2010hb}. In the absence of a first-principles determination, it is therefore reasonable to adopt $\xi=0.151$, fixed in the light sector,\,\cite{Dang:2023ysl}, as an upper bound. Accordingly, in Fig.\,\ref{fig:xi} we display the $\xi$-dependence of the decay width $\Gamma[\eta_c\to\gamma\gamma]$, including a standard error arising from the range considered for the ultraviolet cutoff $\Lambda_{uv}$. Unsurprisingly, the decay width grows as the strength of the AMM increases\,\cite{Sultan:2024mva}.

\begin{figure}[!htbp]
 \centering
 \includegraphics[scale=0.6]{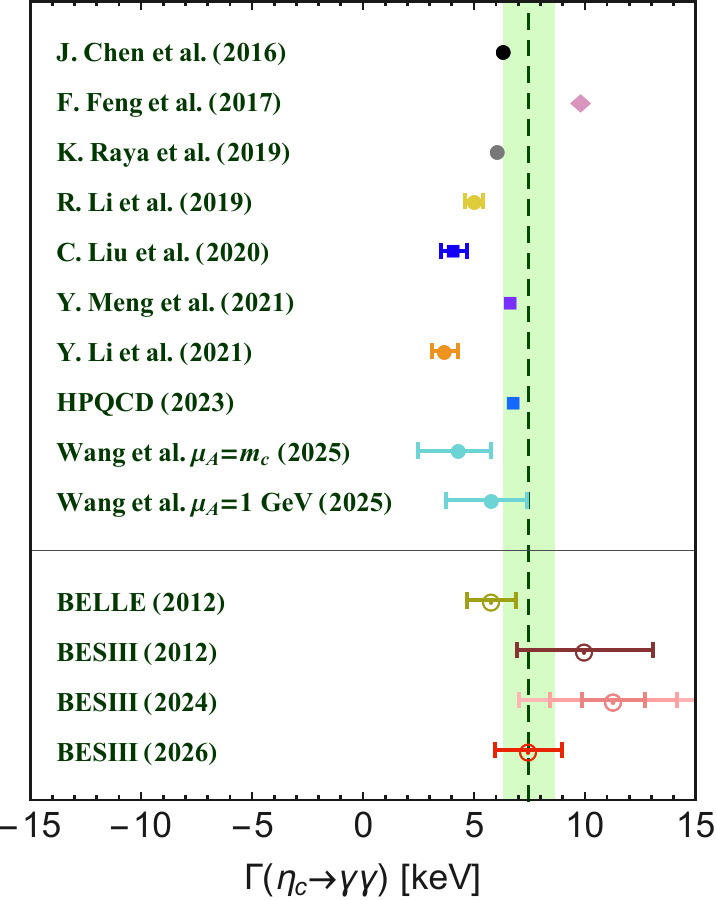}
   \caption{$\Gamma[\eta_c\to\gamma\gamma]$ compared with other theoretical predictions and experimental measurements~\cite{Feng:2017hlu,Chen:2016bpj,Raya:2019dnh,Li:2019ncs,Li:2021ejv,Wang:2025afy,Liu:2020qfz,Colquhoun:2023zbc,Meng:2021ecs,Belle:2012uhr,BESIII:2012lxx,BESIII:2024rex,BESIII:2026pff}. Our result is indicated by a vertical dashed line with its corresponding uncertainty band.}
 
   \label{fig:comparison}
\end{figure}

A comparison of our results for $\Gamma[\eta_c\to\gamma\gamma]$ with many theoretical approaches and phenomenological extractions is presented in Fig.\,\ref{fig:comparison}. Our assesment shows clear compatibility with a range of theoretical predictions, including other CSM studies with more sophisticated numerical implementations\,\cite{Chen:2016bpj,Raya:2019dnh}, lQCD evaluations\,\cite{Meng:2021ecs,Colquhoun:2023zbc}, and the most recent 2026 BESIII extraction\,\cite{BESIII:2026pff}, which aligns with the world-average\,\cite{ParticleDataGroup:2024cfk}. Despite this fact and even though our results are significantly lower than the 2024 BESIII measurement\,\cite{BESIII:2024rex}, this cannot be conclusively ruled out at the present level of sophistication.

\section{Summary}\label{sec:summary}
In this work, we have examined the $\eta_c\to\gamma\gamma$ and $J/\psi\to\gamma\eta_c$ radiative decays within a symmetry-preserving CI model. The present framework extends previous investigations of these transitions within the CI model by systematically and dynamically incorporating the effects of the quark anomalous magnetic moment. To date, such effects have also not been properly disentangled in QCD-connected approaches with momentum-dependent interactions. Our main objective was to examine whether the AMM contribution helps reconcile the results with the 2024 BESIII measurement of the decay width $\Gamma[\eta_c\to\gamma\gamma]$, which lies above typical values, as well as with the most recent 2026 BESIII extraction.

Our study shows that, while quark AMM effects can indeed increase the decay widths under consideration within a reasonable parameter range, they are not sufficient on their own to explain the 2024 BESIII results. Nevertheless, the inclusion of the AMM does lead to improved agreement with  phenomenological expectations and lQCD results for the $\eta_c\to\gamma\gamma$ and $J/\psi\to\gamma\eta_c$ decays. In this sense, the framework naturally accommodates the 2026 BESIII results as well as the current world average.

At the present level of precision, this outcome does not allow a definitive judgment on the validity of either measurement. What it does make clear, however, is the presence of a tension: the 2024 value falls outside the range naturally described by the CI+AMM framework, while the 2026 result is well accommodated. This suggests that the situation is not yet settled, and that the 2024 measurement may benefit from further experimental scrutiny. On the theory side, more refined approaches would also be valuable to improve the picture.

\section*{Acknowledgements}
Work supported by: National Natural Science Foundation of China (Grant No. 12135007); Postdoctoral Fellowship Program of CPSF under Grant Number GZC20240759; Spanish Ministry of Science and Innovation (MICINN grant no.\ PID2022-140440NB-C22); and Junta de Andalucía (grant no.\ P18-FR-5057).

\appendix

\section{Transition form factor}
\label{ap:TFFs}

\begin{figure}[!htbp]
 \centering
 \includegraphics[scale=0.69]{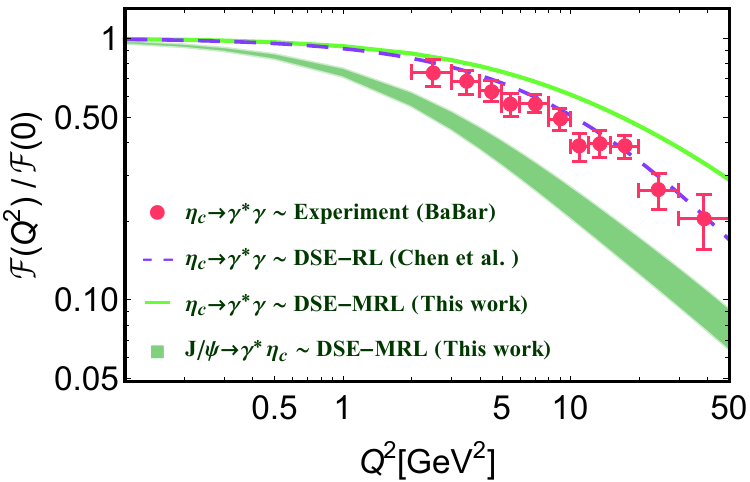}
   \caption{Two-photon transition form factor of the $\eta_c$ meson, compared to a momentum-dependent RL evaluation,\,\cite{Chen:2016bpj}, and experimental data from BaBar,\,\cite{BaBar:2010siw}. The $J/\psi\to\gamma^*\eta_c$ is also included.}
   \label{fig:FF}
\end{figure}

As a supplement to the main text, we provide the transition form factor (TFF). Our result of $\eta_c\to\gamma^*\gamma$ is shown in Fig.\,\ref{fig:FF}. This is compared with a previous CSM assessment,\,\cite{Chen:2016bpj}, and the experimental extraction from BaBar\,\cite{BaBar:2010siw}. Owing to the momentum-independent nature of the CI model, the $Q^2$-dependence of the form factor shows a milder falloff. Nevertheless, the difference between our momentum-independent result and the momentum-dependent one,\,\cite{Chen:2016bpj}, is not as pronounced as one finds for the pion case (see e.g.\,\cite{Dang:2023ysl}). The $J/\psi\to\gamma^*\eta_c$ TFF is also included in Fig.\,\ref{fig:FF}. This exhibits a steeper falloff compared $\eta_c\to\gamma^*\gamma$, as is likewise observed in analogous processes involving the $\rho-\pi$ mesons\,\cite{Maris:2002mz}.

A typical way to quantify the hardness of TFFs is through the interaction radius, defined for a given form factor $\mathcal{F}$ as follows:
\begin{equation}
    r_{\mathcal{F}}^2 = -\frac{6}{\mathcal{F}(0)}\frac{\partial \mathcal{F}(Q^2)}{\partial Q^2}\Bigg|_{Q^2=0}\,.
\end{equation}
Herein, we obtain $r_{\eta_c\gamma}=0.133(1)\,\text{fm}$ and $r_{J/\psi\to\eta_c}=0.28(1)\,\text{fm}$. The $r_{\eta_c\gamma}$ radius lies below estimates from momentum-dependent RL computations\,\cite{Raya:2016yuj}, lQCD evaluations\,\cite{Dudek:2007zz}, and phenomenological extractions\,\cite{Druzhinin:2010zza}; respectively: $r_{\eta_c\gamma}\sim0.16,\,0.14,\,0.17\,\text{fm}$. The $r_{J/\psi\to\eta_c}$ radius is likewise expected to be smaller than that obtained within a momentum-dependent interaction. 

Notwithstanding these differences in the absolute scale, the ratio $r_{\eta_c\gamma}/r_{\pi\gamma}\approx 1/5$ is preserved, provided $r_{\pi\gamma}$ is computed consistently within the same framework (in our case, within the CI model\,\cite{Dang:2023ysl}). This suggests that the CI effectively induces an overall rescaling of radii while preserving the relative hierarchy between light- and heavy-quark systems.

In any case, a comprehensive momentum-dependent treatment with explicitly identifiable AMM effects is not yet available.

\bibliography{ref}

\end{document}